\documentclass[aps,prl,twocolumn,superscriptaddress]{revtex4-1}
\usepackage{graphicx}
\usepackage[rgb,dvipsnames]{xcolor}
\usepackage{amsmath}
\usepackage{amsthm, amssymb}
\usepackage{epsfig}
\usepackage{hyperref}
\usepackage{bm}

\usepackage{natbib}

\usepackage[normalem]{ulem}

\usepackage{placeins}
\usepackage[symbol]{footmisc}

\usepackage{tikz}
\usetikzlibrary{trees}
\begin{document}


\title{Mechanics promotes coherence in heterogeneous active media}

\author{Soling Zimik}
\affiliation{The Institute of Mathematical Sciences, CIT Campus, Taramani, Chennai 600113, India}

\author{Sitabhra Sinha}
\affiliation{The Institute of Mathematical Sciences, CIT Campus, Taramani, Chennai 600113, India}
\affiliation{Homi Bhabha National Institute, Anushaktinagar, Mumbai 400094, India}

\date{\today}

\begin{abstract}
Synchronization of activity among myocytes constituting vital organs, e.g., the heart, is crucial for physiological functions. Self-organized coordination in such heterogeneous ensemble of excitable and oscillatory cells is therefore of clinical importance. We show by varying the strength of intercellular coupling and the electrophysiological diversity, a wide range of collective behavior emerges including clusters of synchronized activity. Strikingly, stretch-activated currents allow waves of mechanical deformation to alter the activity of neighboring cells, promoting robust global coherence.
\end{abstract}

%
%

\maketitle

Cells whose contractility are mediated by electrical signals are central to many vital physiological processes~\cite{Bers2002,Langton1989,Zheng2018}.
Examples include
the myocytes in the heart and smooth muscle cells in the uterus and the
gastro-intestinal tract\cite{Pfeiffer2014,Young2016,Huizinga2009}. While the mechanisms that result in changes in
the electrical state of the cells (in the form of action
potentials) give rise to contractions of the myofibrils - a process
referred to as excitation-contraction coupling - has been investigated 
at great depth, the manifestation - at the tissue-scale - of communication between 
neighboring cells through such interacting electrical and mechanical activities
is less well-understood. 
The problem is further compounded by the fact that most biological tissues are
heterogeneous mixtures of different cell types having distinct characteristic
patterns of local activity - which makes the coherence
achieved by the system even more remarkable.
As intercellular interactions are believed to underlie the observed system-wide 
synchronization
of activity that is functionally crucial, e.g., as in the pumping action
of the heart, it is important to arrive at a comprehensive explanation
of how global coordination can be achieved by these means.

\begin{figure}[tbp]
    \centering
        \includegraphics[scale=0.6]{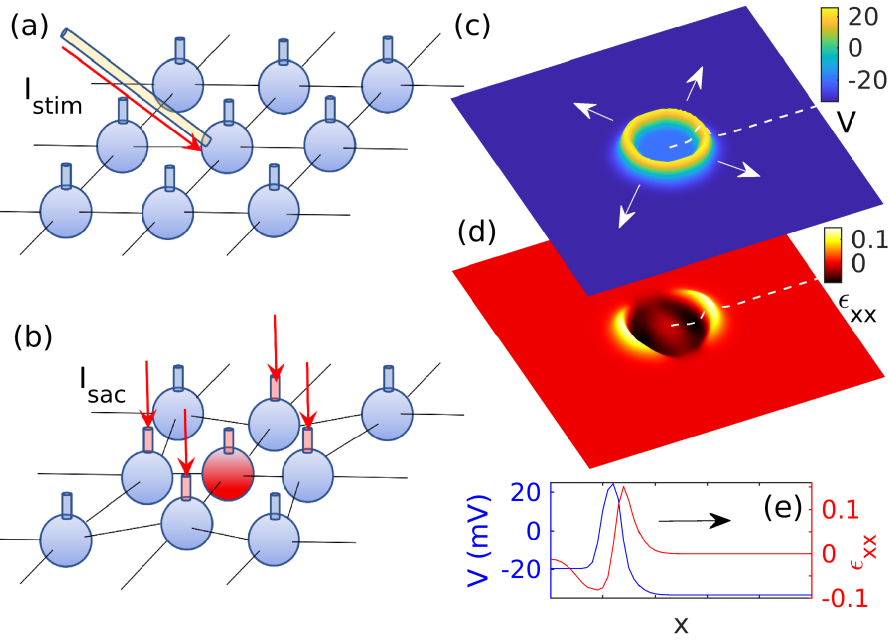}
\caption{Inter-cellular electromechanical coupling promotes propagation
of activity across excitable media.
(a-b) Excitable tissue schematically represented as a regular lattice of
cells, each coupled to its neighbors both electrically and
mechanically. (a) Activation of the central cell upon stimulation by
current $I_{stim}$ results in a contraction that (b) locally
deforms the neighborhood and induces activity in surrounding cells
by allowing current $I_{sac}$ through their stretch-activated channels.
(c-d) Point stimulation at the center of a domain results in
propagation (indicated by arrows) of an excitation front in the 
membrane potential $V$ (c), accompanied by a deformation manifesting as a
wave in the normal strain field, whose $x$ component, $\epsilon_{xx}$,
is shown in (d).
The spatial profiles of the two waves (blue: $V$, red: $\epsilon_{xx}$) 
along the broken line are indicated in (e). Note the deformation
at locations anterior to the excitation front, moving in the
direction of the arrow, that results from stretching
($+$ve strain)
induced in its neighbors by a cell contracting ($-$ve strain)
upon being excited.}
\label{fig1}
\end{figure}

Cells responding to electromechanical signals typically possess
the property of \textit{excitability}~\cite{meron1992pattern,sinha2014patterns,Nitsan2016,Jia2023}, characterized by a nonlinear response of the transmembrane potential to
electrical currents. In excitable ($E$) cells, this manifests
as an action potential elicited by a supra-threshold stimulus. Other cells,
characterized as oscillatory ($O$) cells, spontaneously generate
periodic pulses of activity. These distinct cell types coexist in many
natural settings, such as, 
the GI tract~\cite{Sanders2023}, the sino-atrial node, i.e., the natural pacemaker of the heart~\cite{camelliti2004fibroblast} or diseased cardiac tissue~\cite{nguyen2014cardiac,Okabe2024}.
Note that $O$ cells can even represent cellular aggregates comprising excitable and passive cells 
(e.g., fibroblasts) coupled
by gap junctions that can give rise to emergent oscillations~\cite{Jacquemet,singh2012self}. 
These are of relevance, e.g., in the uterus, where single, isolated cells cannot exhibit autonomous activity but sites of spontaneous activity nevertheless
arise during the later stages of pregnancy to coordinate organ-wide contractions
that are essential for childbirth~\cite{Young2016}.
As the relative density of such cells and the strength of their coupling can 
change over time, for instance,
in the embryonic heart~\cite{watanabe2016probing,coppen2003comparison}, in the
gravid uterus~\cite{lenhart1999expression,Young2016,risek1991spatiotemporal}, 
or as a result of disease-related physiological alterations~\cite{severs2004gap,nguyen2014cardiac}, it is essential to explicate the 
collective dynamics resulting from electrical and mechanical interactions between these heterogeneous cell types.
In particular, one would like to understand the mechanisms that promote robust
coherence of periodic activity that is crucial for the functioning
of these organs and whose breakdown lead to pathological outcomes such
as cardiac arrhythmia, irritable bowel syndrome or premature contractions
leading to pre-term birth.
A crucial piece of the puzzle that has become manifest only relatively recently
is provided by the coupling between electrical excitation of a cell and the
resulting mechanical contraction~\cite{Weise2012,Nitsan2016}. This stress in turn affects
the activation of neighboring cells via mechanosensitive pathways, 
e.g., by inducing stretch-activated current ($I_{sac}$) through specialized channels [Fig.~\ref{fig1}~(a-b)]. 
Thus, a wave of excitation propagating radially outward that is initiated by a local
stimulation will be preceded by a mechanical deformation (stretching) 
of cells anterior
to the wavefront, thereby stimulating $I_{sac}$ (and hence excitation)
in these cells [Fig.~\ref{fig1}~(c-e)].

In this paper we show that such electromechanical interactions between cells
enhances oscillatory activity in the system significantly over and above the
contribution from electrotonic coupling via gap junctions. Indeed, even for
situations where such electrical interactions are weak, the
cells can coordinate by means of the tension-induced
electrical activity, resulting in the emergence of robust coherence, 
with almost all cells oscillating in phase.
We demonstrate the key role played by mechanics in promoting
synchrony in the context of electrically heterogeneous media
comprising both $E$ and $O$ cells~\cite{bub2002spiral,bub2005global,kryukov2008synchronization,singh2012self}. 
As the relative density
of the distinct cell types and/or the strengths of electrotonic coupling
between them are varied, their collective activity exhibits transitions
between a variety of spatiotemporal patterns, e.g., spatially localized 
clusters of excitation eventually merging to states characterized by
traveling waves, ultimately leading to coherence.
As these in turn can be associated with various abnormal physiological rhythms~\cite{glukhov2013sinoatrial,kharche2017computational}, our
results can have potential implications of clinical relevance.   

\begin{figure}[htbp!]
    \centering
        \includegraphics[scale=0.6]{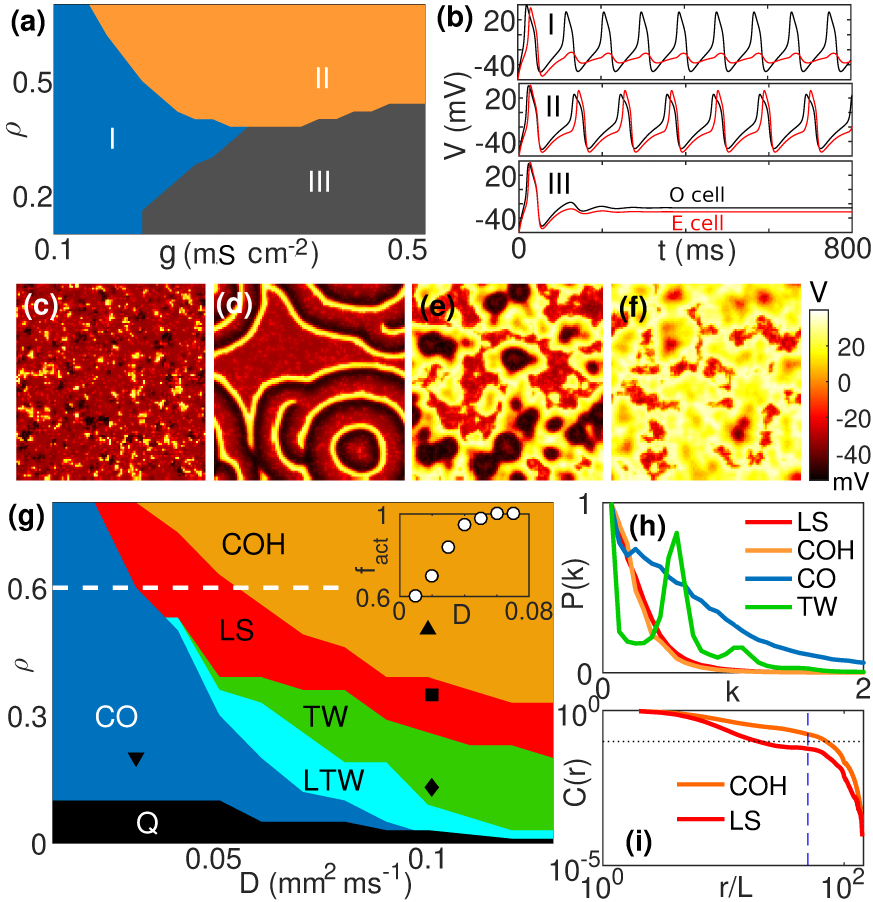}
\caption{
Heterogeneous media, comprising excitable ($E$) and oscillatory ($O$) cells
that are electrically coupled to neighbors, can display diverse
spatiotemporal patterns, with coherent activity emerging only
under strong coupling and for relatively higher proportion of oscillatory 
cells.
(a-b) The collective dynamical regimes observed in the mean-field limit 
for different values of coupling, $D$, and relative fraction 
of $O$ cells, $\rho$, correspond to: (I) $O$ active, $E$ inactive,
(II) both $O$ and $E$ active, and (III) both $O$ and $E$ inactive
[$O$ and $E$ are indicated by red and blue curves, respectively,
in the time-series shown in (b)].
(c-f) Different spatiotemporal activity patterns observed in $V$ for a
$2$-dimensional square lattice ($L=50$) with $\rho$ fraction of $O$ and $(1-\rho$) fraction
of $E$ cells, neighbors being coupled with strength $D$: (c) Clustered
Oscillation (CO), 
(d) Traveling Waves (TW),
(e) Localized Synchronization (LS),
and (f) Coherence (COH) [Video S1].
(g) Oscillatory activity is suppressed upon decreasing $\rho$, yielding
Quiescent state Q.
Localized Traveling Waves (LTW) regime may also be observed, in which
the fraction of active cells $f_{act}<0.9$ [Video S2].
The inset shows the variation of $f_{act}$ with $D$ along the broken line 
in the main figure to indicate the distinction between CO and LS
regimes.
The parameter values for the patterns in (c)-(f)
are indicated using the markers $\bigtriangledown$, $\Diamond$, $\square$ and 
$\triangle$, respectively. The corresponding states 
can be distinguished by their
(h) power spectrum $P(k)$ and 
(i) correlation function $C(r)$ [horizontal dotted line: $C(r)=0.1$; vertical
broken line: $r = L/2$, $L$ being the linear extent of the domain].
}
\label{fig2}
\end{figure}

The time-evolution of the state (characterized by the transmembrane potential $V$)
of the tissue considered as a continuum is given by
\begin{equation}
\partial V/\partial t = (I / C_m) + \nabla . ( D \nabla V) - \nabla . (V \dot{\textbf{U}}),
\label{eqn1}
\end{equation}
where $C_m$ ($=20~\mu$F cm$^{-2}$) is the membrane capacitance, and the current comprises $I = I_{bas} - I_{ion} + I_{sac}$. Here, $I_{bas}$ is a basal current that determines whether an isolated cell is 
capable of spontaneous oscillations (for $I_{bas}>90$, via a Hopf 
bifurcation~\cite{tsumoto2006bifurcations})
and $D$ is the diffusion coefficient representing the gap-junctional conductances that can vary across space.
The ionic current for each cell is described according to Morris-Lecar equations~\cite{morris1981voltage}
and comprises a $K^{+}$ current determining
the recovery, a 
$Ca^{2+}$ current providing the initial excitation, and a membrane leakage 
current, viz., 
$I_{ion} = g_K n (V - E_K) + g_{Ca} m_{\infty} (V - E_{Ca}) + g_L (V - E_L)$,
where $n$ is a gating variable for the K$^+$ current whose time-evolution is described by $dn/dt = (n_{\infty} - n)/\tau_n$. The kinetics of the $Ca^{2+}$ channel is assumed to be relatively faster and thus, the asymptotic value
of $m_{\infty} = [1 + \tanh(V-V_1)/V_2]/2$ is used. The parameters are
chosen to be $g_L = 2$, $g_{Ca} =4.4$, $g_K = 8$ (m$\mho$ cm$^{-2}$), $V_1 = -1.2$, 
$V_2 = 18$, $E_K = -84$, $E_{Ca}=120$ and $E_L = -60$ (mV).
The recovery of the cell after activation is governed by the $K^{+}$ current, which, in turn, is determined by the gating variable $n$ 
with its asymptotic value $n_{\infty} = [1 + \tanh(V-V_3)/V_4]/2$ and the
characteristic time $\tau_n = 25/\cosh([V-V_3]/2 V_4)$, the parameters being
set to $V_3 = 2$, $V_4 = 30$~\cite{Ermentrout2010}. 
The stretch-activated current $I_{sac} = g_{sac}~E~(V-V_{sac})$ represents the electrical
contribution of mechanical interactions between the cells, with $g_{sac}$ and 
$V_{sac}$ representing conductance and reversal potential of the corresponding
channel~\cite{kuijpers2007mechanoelectric,Weise2017,note2}. 
As stretching of
the cell results in the opening of the channel~\cite{SimonChica2024,Reed2014}, 
$I_{sac}$ is finite if the normal components of the strain $\epsilon$ are positive, measured as
$E=\sqrt{\smash[b]{\epsilon_{xx}^2\Theta(\epsilon_{xx})+\epsilon_{yy}^2\Theta(\epsilon_{yy})})}$,
where $\epsilon_{ij} = \tfrac{1}{2}(\partial U_i/\partial r_j + \partial U_j/\partial r_i)$  with $\{i,j\}=\{x,y\}$,
$U_i$ are the components of the displacement 
vector field {\bf U}, $\boldsymbol{r}$ is the position vector, and $\Theta (~)$ is the Heaviside step function.
We assume the medium to be a linear isotropic elastic solid in an overdamped 
environment~\cite{yuval2013dynamics,note1}, 
such that the displacement evolves as
\begin{equation}
\partial{\textbf{U}}/\partial t =\nabla .(\boldsymbol{\sigma}^{passive} + \boldsymbol{\sigma}^{active}),
\label{eqn3}
\end{equation}
where the passive component of the stress is 
determined by the material properties of the solid, viz.,
$\sigma^{passive}_{ij} = \lambda \delta_{ij},
\epsilon_{kk}+2 \mu \epsilon_{ij}$, where
$\delta_{ij}$ is the Kronecker delta function and $\lambda, \mu$ are the Lam\'{e} parameters~\cite{note4}.
The active component $\sigma^{active}_{ij} = T^{active}~\delta_{ij}$ results from the
active tension $T^{active}$ induced by intrinsic electrical excitation of a cell and evolves
with the normalized transmembrane potential $\mathcal{V}$ [$=(V - V_{\rm min})/(V_{\rm max} - V_{\rm min})$, where $V_{\rm max}=30$ mV and $V_{\rm max}=-50$ mV are the highest and lowest
$V$ values attained during an action potential] as:
\begin{equation}
    \frac{dT^{active}}{dt} = K (\alpha \mathcal{V}-T^{active}),
\label{eqn4}
\end{equation}
where $K(\mathcal{V})$ is a function that controls the time-scale of active stress, which is defined by $K(\mathcal{V})= 0.02 (0.2)$ for $\mathcal{V}<-0.01$ ($\mathcal{V}>-0.01$) and $\alpha (=0.3)$ controls the magnitude of the contraction
impulse~\cite{nash2004electromechanical}.
Thus, Eqns.~(\ref{eqn1}-\ref{eqn4}) model the electromechanical 
coupling in the medium.
The activation (rise in $V$) of a cell increases the tension $T^{active}$ 
inside the cell, as in Eqn.~(\ref{eqn4}), causing its contraction, which in turn 
stretches its neighboring cells and activates $I_{sac}$ in the neighboring 
cells. The final (convective) term of Eqn.~(\ref{eqn1}) represents the contribution to the change in potential
resulting from transport due to tissue deformation. 

\begin{figure}
    \centering
        \includegraphics[scale=0.6]{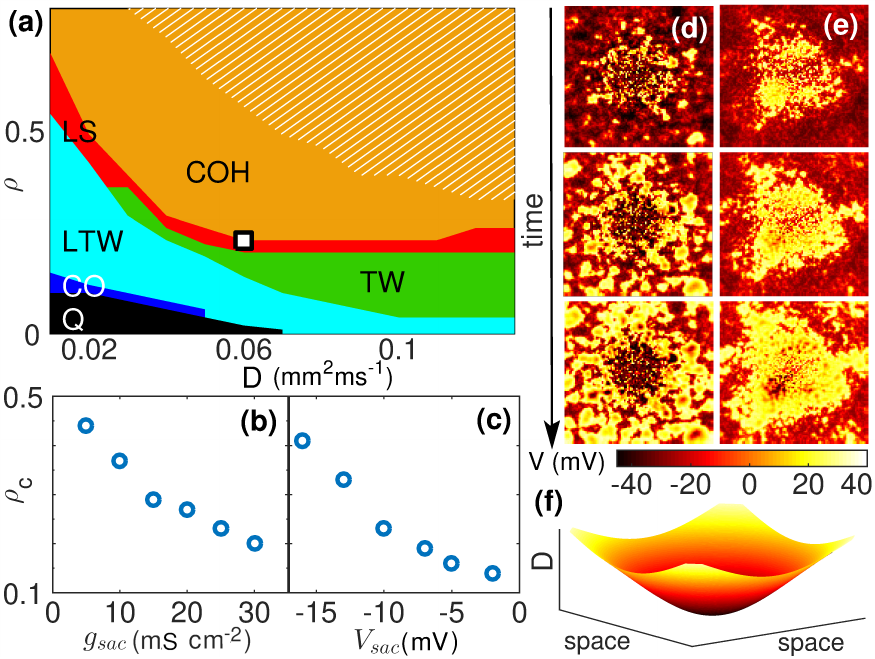}
\caption{Mechanical interactions promote synchronization of activity
even for extremely weak electrical coupling between cells.
(a) In presence of $I_{sac}$ ($g_{sac} = 25, V_{sac}=-10$), coherence (COH) is observed
over a much broader range of parameters $\rho$ and $D$ [for comparison,
the hatched region shows its extent in the absence of $I_{sac}$, see
Fig.~\ref{fig2}~(g)]. The other globally oscillating states ($f_{act}>0.9$),
LS and TW, 
have extended into the regions that was occupied by
the quiescent regime Q and the partially oscillatory CO domain
when $g_{sac} =0$ [see Fig.~\ref{fig2}~(g)].
The critical value $\rho_c$ of oscillator density 
at which COH emerges for $D=0.06$ is indicated (open square).
(b-c) Variation of $\rho_c$ as a function of the conductance $g_{sac}$ [$V_{sac} = -10$] (b)
and reversal potential $V_{sac}$ [$g_{sac} = 25$] (c) of the stretch activated channel, 
respectively, implying that coherence can be achieved using relatively
fewer $O$ cells by increasing either of these channel parameters.
(d-e) Snapshots indicating the spatiotemporal evolution of $V$ (time increasing from top to bottom)
for $g_{sac} = 0$~(d) and $25$~(e), respectively. The spatially changing strength
of electrical gap-junction couplings is modeled by $D$ varying along 
a gradient between [$0,0.07$], increasing radially from the center (f). 
Comparing (e) with (d), we note that the
activity becomes more coherent (particularly in the central region) 
in the presence of mechanical interactions [Video S3].
}   
\label{fig3}
\end{figure}

We consider a square lattice of $L \times L$ cells ($L=100$ for the simulations
shown here, although we have verified that qualitatively similar results
can be seen with other system sizes). Of these, a fraction $\rho$ are 
oscillatory ($O$) cells capable
of spontaneous periodic activation, the remaining cells being excitable ($E$). The former
have $I_{bas}^{O}$ that are randomly sampled from an uniform distribution bounded
between $[90,150]$, while the latter are characterized by $I_{bas}^{E} = 70$.
Each cell is coupled diffusively to its neighboring four cells in the square lattice for the simulations reported here.
The diffusive coupling strength $D$ is in general a function of space, but is taken
to be a constant for most of our simulations, except those shown in
Fig.~\ref{fig3}~(d-e), where it has a spatial gradient.
The PDEs (\ref{eqn1}) and (\ref{eqn3}) are solved using a pseudospectral method \cite{Canuto1988}
for the spatial part (spatial resolution $\delta{x}=2$ mm) with periodic boundary conditions, while the temporal evolution of all the equations are performed by the 
Euler method with a temporal resolution $\delta{t}=0.1$~ms. 
For each choice of $\rho$ and $D$, results have been averaged over $10$
realizations. 
%

We begin by exploring the conditions under which the model exhibits
collective dynamics characterized by various spatiotemporal activation patterns,
in the absence of any mechanical interactions between the cells. To get a broad understanding
of the principal regimes that are possible, we can simplify further to obtain a mean-field 
description of the medium in which each cell has exactly the same fraction $\rho$
and $1 - \rho$ of $O$ and $E$ cells, respectively, among its $k$ neighbors ($k=4$ for
the lattices we have shown results for). We note that such a setting will give results
identical to the spatially heterogeneous situation implicit in Eqn.~(\ref{eqn1}) 
in the limit of extremely strong gap-junctional coupling (represented by $D \rightarrow \infty$) when the system can be
effectively reduced to a mutually coupled pair of $O$ and $E$ elements whose
membrane potentials, $V^O$ and $V^E$, respectively, evolve as
$C_m d{V^O}/d{t}=-I_{ion} + I_{bas}^{O}+ k g (1-\rho)(V^E-V^O)$ and 
$C_m d{V^E}/d{t}=-I_{ion} + I_{bas}^{E} + k g \rho(V^O-V^E)$. 
As seen in Fig.~\ref{fig2}~(a), in this mean-field setting, the system can be in any one
of three regimes depending on the density of $O$ cells $\rho$ and the intercellular 
coupling strength $g$, viz., I: only the $O$ cells show appreciable
activity (for low $g$), II: $O$ and $E$ cells oscillate in phase (for high $\rho$ and $g$)
and III: both cells inactive (for low $\rho$ but high $g$). These regimes can be
intuitively understood as follows: in regime I, the coupling is too weak for the $O$ cells 
(that exhibit spontaneous activity) to drive the $E$ cells above their threshold - a situation
that is rectified upon making the coupling stronger (regime II). However, if the density
of $O$ cells is relatively low (regime III), increased coupling leads to a source-sink imbalance such
that the gap-junctional current from $O$ cells is shared between too many inactive 
cells and is insufficient to drive any of the cells over the threshold,
resulting in both cell types becoming quiescent.

Collective dynamics analogous to these three regimes can be observed
even upon 
introducing spatial heterogeneity in the domain by randomly choosing
the fraction of oscillatory neighbors of a cell
from a distribution having a mean value of $\rho$.
Thus, the clustered oscillations (CO) state obtained in the limit of weak coupling
[Fig.~\ref{fig2}~(c)], in which the $O$ cells are active but 
are unable to drive all the $E$ cells because of the low value of $D$, 
is analogous to Regime I in Fig.~\ref{fig2}~(a). 
Note though that the heterogeneity introduces a degree of variability
absent in the mean-field case, viz., instead of only the $O$
cells being active at low $D$, activity is seen across small spatial clusters
that locally have a high density of $O$ cells. 
Increasing the coupling strength results in differential outcomes
depending on $\rho$. For a low density of $O$ cells, this leads to
cells becoming quiescent (Q) because of source-sink mismatch,
analogous to the situation for regime III in the mean-field limit.
As $\rho$ is increased, the relatively strong coupling between cells
imply that local clusters having a higher density of $O$
cells act as organizing centers for activity spreading through the
domain in the form of traveling waves [TW, Fig.~\ref{fig2}~(d)].
As $\rho$ is increased further, we observe successively larger
regions of the domain getting synchronized in their activity in
the locally synchronized (LS) regime [Fig.~\ref{fig2}~(e)], eventually
leading to coherence [COH, Fig.~\ref{fig2}~(f)], such that the periodic
activity of cells across the entire domain is occurring in phase, analogous
to regime II in the mean-field limit.

The various types of collective dynamics discussed above arise under
different conditions of cellular heterogeneity 
[quantifiable as $\sim \rho ( 1 - \rho)$]
and coupling, as seen from the position of the corresponding domains 
in $\rho$-$D$ parameter space shown in Fig.~\ref{fig2}~(g).
The inset shows the variation with $D$ in the order parameter $f_{act}$, i.e.,
the fraction of cells undergoing persistent oscillations, for
maximal heterogeneity in the cellular composition of the domain
($\rho = 0.5$). A transition from CO to LS can be observed when
$f_{act}$ exceeds $90 \%$ around $D \sim 0.03$.
As already mentioned, if the relative density of $O$ cells
are decreased, then depending on the coupling strength between cells, 
we will either see a transition to Q (when $f_{act} <0.1$) 
or TW. The latter pattern is of particular relevance for systems
such as the sino-atrial node or uterine tissue where they have
been observed~\cite{glukhov2013sinoatrial,Lammers2008} and
associated with aberrant functioning~\cite{kharche2017computational,Xu2015}.
The TW regime can be identified by the presence of a characteristic
wavelength in the resulting patterns, resulting in a prominent secondary peak
[Fig.~\ref{fig2}~(h)]
in the power spectrum $P(k)$ ($=\tilde{V}_k \tilde{V}_k^{\ast}$, where
$\tilde{V}_k, \tilde{V}_k^{\ast}$ are the amplitudes of the $k^{\rm th}$ Fourier 
mode of $V(r)$ and its complex conjugate).
If we decrease the density of $O$ cells, the waves are eventually confined within
spatially disjoint patches ($0.1 < f_{act} <0.9$), which we refer
to as \textit{Localized} Traveling Waves (LTW) [Video S2]. 

As reported above, in the strong coupling regime, increasing the fraction
of $O$ cells results in the clusters exhibiting 
spontaneous activity becoming larger in size.
The increased degree of synchronization seen as the system traverses
through the LS and COH regimes with increasing $\rho$ is reflected in
the reduced width of the distribution of mode amplitudes $P(k)$ 
[Fig.~\ref{fig2}~(h)]. This suggests that activity in the system is
coordinated over larger length scales (corresponding to small $k$),
which is shown explicitly by measuring the spatial correlation function
$C(r)=\langle V(r_0)V(r_0+r)\rangle -\langle V(r_0)\rangle^{2}/ (\langle V(r_0)^{2}\rangle-\langle V(r_0)\rangle^{2})$ between the activity at a pair of
locations separated by a distance $r$, the averaging being
over both $r_0$ and time. 
The COH regime is distinguished from LS by the
correlation length $\xi$ over which
$C(r)$ decreases to $10\%$ of its maximum value [Fig.~\ref{fig2}~(i)].
COH is characterized by $\xi \ge L/2$ ($L$ being the system size),
implying that almost all points in the medium are correlated.

On introducing the contribution of mechanical interactions between cells 
to the electrical activity via $I_{sac}$, which is modulated by tissue deformation (see above), we observe that 
synchronization is promoted in the system. This is evident from the expanded
region in the $\rho$-$D$ parameter space corresponding to COH 
[Fig.~\ref{fig3}~(a)] in comparison to Fig~\ref{fig2}~(a), as well as,
the occurrence of  LS, TW and LTW even with extremely weak gap-junctional coupling.
The stretch-activated current results in the system being capable of 
spontaneous periodic activity over almost the entire parameter space,
indicated by the significant shrinking of CO and Q.
The role of the stretch activated channel parameters, specifically, 
$g_{sac}$ and $V_{sac}$, in driving COH can be seen from the decrease 
in the critical value
$\rho_c$, the minimum relative density of $O$ cells with which coherence
can be
achieved (for $D=0.06$), as the channel conductance and reversal potential
are increased, respectively [Fig.~\ref{fig3}~(b-c)].

Mechanical interactions between the cells can also enhance synchronization
in tissue where the electrical coupling displays spatial heterogeneity,
e.g., a spatial gradient in the effective gap-junction conductance as
has been reported in biological excitable systems such as the sino-atrial 
node in the heart~\cite{verheijck2001electrophysiological,honjo2002heterogeneous}.
Figs.~\ref{fig3}~(d-e) compares the spatiotemporal evolution
of activity in the absence and presence, respectively, of mechanical 
interactions, in a medium where the electrical coupling $D$ between cells
increases radially from the center [as shown in Fig.~\ref{fig3}~(f)]. 
At each cycle of periodic excitation, the activity begins in the
central region that has the weakest coupling which promotes stimulation
by reducing source-sink mismatch between $O$ and $E$ cells.
Without the stretch-activated current, however, as this activity spreads
outward from the center it gets fragmented because the low density
of gap junctions prevent many of the cells from getting excited
simultaneously [Fig.~\ref{fig3}~(d)]. The introduction of $I_{sac}$ provides
an alternative mechanism by which cells can excite their neighbors,
resulting in a more coherent pattern of spreading activity 
[Fig.~\ref{fig3}~(e) and Video S3].

To conclude, we show that mechanical interactions between cells
can work in tandem with electrical intercellular communication 
via gap junctions to 
create a robust mechanism for promoting synchronized
oscillations in heterogeneous cell assemblies. Indeed, in the limit
of weak electrotonic coupling, communication mediated by the physical
deformation induced by excitation could be the primary means
of coordinating collective behavior of cellular arrays.
Our results are consistent with, and may help explain, recent 
experiments reporting the crucial role that mechanical interactions 
play in coordinating activity, e.g., in the embryonic heart~\cite{chiou2016mechanical}. 
Our observation of diverse spatiotemporal patterns can be of
relevance in understanding emergent collective dynamics in 
media where $\rho$ and $D$
can vary across space (forming a gradient, as that of gap-junctional
density in the sino-atrial node~\cite{verheijck2001electrophysiological,honjo2002heterogeneous}) 
and/or in time under natural or pathological settings (such as, the changes
observed in the
developing embryonic heart \cite{watanabe2016probing,coppen2003comparison},
the gravid uterus~\cite{lenhart1999expression,Young2016,risek1991spatiotemporal} 
and progressively fibrotic cardiac
tissue~\cite{severs2004gap,nguyen2014cardiac}).
The coherence enhancing mechanism described here is based on a 
very generic model, suggesting that it may be valid for 
a broad class of systems~\cite{Jia2023,Lammers2008,Huizinga2009,Manchanda2019}.
Thus our results may provide insights about the mechanisms 
underlying coordination of collective activity - and hence, understanding
pathologies associated with its breakdown - in heterogeneous tissue
comprising excitable and oscillatory cells, as in the sino-atrial node,
the gastro-intestinal tract and the gravid uterus.
%

\begin{acknowledgments}
SZ has been supported by the Center of Excellence in Complex Systems and Data 
Science, funded by the Department of Atomic Energy, Government of 
India. We would like thank Shakti N Menon for helpful suggestions.
\end{acknowledgments}


\begin{thebibliography}{47}
\bibitem{note4} The numerical values for the Lam\'{e} parameters used in the simulations reported here are $\lambda=\mu=0.4$ kg mm$^{-1}$ ms$^{-2}$.
\end{thebibliography}
%
\clearpage
\onecolumngrid

\setcounter{figure}{0}
\renewcommand\thefigure{S\arabic{figure}}
\renewcommand\thetable{S\arabic{table}}

\vspace{1cm}
\begin{center}
\textbf{\large{SUPPLEMENTARY INFORMATION}}\\

\vspace{0.5cm}
\textbf{\large{Mechanics promotes coherence in heterogeneous active media}}\\
\vspace{0.5cm}
\textbf{Soling Zimik and Sitabhra Sinha}
\end{center}
\section*{List of Supplementary Movies}
\begin{enumerate}
\item Video S1: Time-evolution of activation patterns for the different dynamical regimes: (top left) Clustered Oscillation or CO, (top right) Traveling Waves or TW,
(bottom left) Localized Synchronization or LS, and (bottom right) Coherence or COH. 
\item Video S2: Spatial localization of traveling waves (TW regime) at relatively low density $\rho$ of oscillating cells and low intercellular
coupling strength $D$.
\item Video S3: Mechanics enhances synchronization as can be observed by comparing the spatiotemporal evolution of activity $V$
in a medium having a spatial gradient in the intercellular coupling strength $D$ with the minimum at the center of the domain
[see Fig.~3~(f) in main text].
In absence of mechanical interactions, the activity is disordered in the central region where $D$ is lowest (left panel), while 
incorporating mechanics results in a high degree of synchronized activity (right panel).
\end{enumerate}
\end{document}